\documentclass[traditabstract]{aa} 
\usepackage{graphicx}
\usepackage{epsfig}
\usepackage{txfonts}
\usepackage{natbib}
\usepackage{lscape}
\bibpunct{(}{)}{;}{a}{}{,} 

\def\asec{\ifmmode ^{\prime\prime}\else$^{\prime\prime}$\fi}

\def\msunyr{\mbox{\,${\rm M_{\odot}\, yr^{-1}}$}}

\def\degs{\ifmmode ^{\circ}\else$^{\circ}$\fi}
\def\amin{\ifmmode ^{\prime}\else$^{\prime}$\fi}
\def\asec{\ifmmode ^{\prime\prime}\else$^{\prime\prime}$\fi}

\def\degs{\ifmmode ^{\circ}\else$^{\circ}$\fi}
\def\amin{\ifmmode ^{\prime}\else$^{\prime}$\fi}
\def\cm{\mbox{\,cm}}

\def\cm3{\mbox{\,cm$^{-3}$}}

\def\lsim{\!\!\!\phantom{\le}\smash{\buildrel{}\over
 {\lower2.5dd\hbox{$\buildrel{\lower2dd\hbox{$\displaystyle<$}}\over
                                 \sim$}}}\,\,}
\def\gsim{\!\!\!\phantom{\ge}\smash{\buildrel{}\over
{\lower2.5dd\hbox{$\buildrel{\lower2dd\hbox{$\displaystyle>$}}\over
                               \sim$}}}\,\,}

\begin{document}

\title{Unveiling the radio counterparts of two binary AGN candidates: J1108+0659
and J1131-0204}

\author{M. Bondi \inst{1}
      \and 
      M.A. P\'erez-Torres \inst{2}
      \and 
      E. Piconcelli \inst{3}
      \and
      H. Fu \inst{4}
}     
          
\institute{ Osservatorio di Radioastronomia - INAF, Bologna,  Via P. Gobetti 
101, 40129 Bologna, Italy\\ 
             \email{bondi@ira.inaf.it}
         \and
             Instituto de Astrof\'{\i}sica de Andaluc\'{\i}a - CSIC,
             PO Box 3004, 18008 Granada,  Spain 
         \and
            Osservatorio Astronomico di Roma -INAF, Via Frascati 33,
            00040 Monte Porzio Catone, Roma, Italy 
         \and
           Department of Physics \& Astronomy, University of Iowa,
           Iowa City, IA 52245, USA
         }
\date{}

 \abstract{The sources SDSS\,J113126.08-020459.2 and SDSS\,J110851.04+065901.4 
are two double-peaked [O III] emitting AGN,
identified as candidate binary AGNs by 
optical and near infrared (NIR) observations.
We observed the two sources with high resolution Very Long Baseline 
Interferometry (VLBI) using the European VLBI Network at 5\,GHz, 
reduced VLA observations at three frequencies 
available for one of the sources, and used archival HST observations. 

For the source SDSS\,J113126.08-020459.2,
the VLBI observations detected only one single compact component associated 
with the eastern NIR nucleus.
In SDSS\,J110851.04+065901.4, the VLBI 
observations did not detect any compact components, but the VLA observations
allowed us to identify a possible compact core in the region of the 
north-western optical/NIR nucleus. In this source we find kpc-scale 
extended radio emission that is
spatially coincident to the ultraviolet continuum and to the
extended emission narrow line region. The UV continuum is significantly 
obscured since the amount of extended radio emission
yields a star formation rate of about 110 \msunyr, which is an order of magnitude
larger than implied by the observed ultraviolet emission.

Our analysis confirms the presence of only one AGN in
the two candidate binary AGNs.}






\titlerunning{}

\maketitle
%

\section{Introduction}

According to the $\Lambda$ cold dark matter hierarchical models of galaxy
formation, more massive galaxies are assembled from smaller ones in a series
of minor or major merger events \citep{1972ApJ...178..623T}.
Since most massive galaxies host a central supermassive black hole (SMBH),
a merger between two galaxies nearly always
results in the formation of a merger-remnant galaxy containing two
SMBHs \citep[e.g.][]{1995ARA&A..33..581K}.
If sufficient gas accretes onto both SMBHs, each may be visible as an 
active galactic nucleus (AGN), forming a so-called ``dual'' or ``binary'' AGN.
Gas and stellar dynamical friction aid the shrinking of the binary AGN, 
which is gravitationally bound by the stellar bulges
with a separation $\gsim 1$ kpc.
The duration of this  stage of the binary AGN is highly uncertain and
depends on the efficiency of the drag from the dynamical friction, but 
the system will ultimately shrink to form a pc-scale binary SMBH 
\citep{1980Natur.287..307B, 2001ApJ...563...34M,2004ApJ...606L..17M, 
2004ApJ...607..765E, 2014SSRv..183..189C}.

In recent years,  
moderate resolution optical spectra surveys, such as the Sloan Digital
Sky Survey \citep[SDSS][and references therein]{2010AJ....139.2360S} and
the DEEP2 Galaxy Redshift Survey \citep{2013ApJS..208....5N},
allowed compilation of catalogues of AGNs with double-peaked [OIII]
line emission that were identified as dual/binary SMBH candidates
\citep{2009ApJ...705L..76W, 2010ApJ...716..866S, 2010ApJ...708..427L, 
2011ApJ...735...48S, 2012ApJ...745...67F}. 
Even though AGN/SMBH pairs are a natural consequence of galaxy mergers and their
search  has received great attention, evidence for such
systems is elusive, so only a few confirmed examples have been
found so far \citep[e.g.][]{2001ApJ...549L.155J, 2003ApJ...582L..15K,
2006A&A...453..433H, 2006ApJ...646...49R,
2008MNRAS.386..105B, 2009ApJ...702L..82C, 2010ApJ...714L.271B,
2010ApJ...722L.147P, 2011ApJ...735L..42K, 2011ApJ...740L..44F, 
2013A&A...557L..14G, 2014Natur.511...57D, 2014ApJ...792L...8W}.
A promising  approach to identifying real dual AGN, based on the 
wide-area high resolution VLA radio observations of the SDSS Stripe 82 
field \citep{2011AJ....142....3H}, has been undertaken by \citet{2015ApJ...799...72F}. This method already confirmed four binary AGN 
\citep{2015ApJ...815L...6F}, and would eventually yield a sample of $\sim 80$
confirmed binaries that are bright enough for VLBI observations.

The galaxies \object{SDSS\,J113126.08-020459.2} and \object{SDSS\, J110851.04+065901.4}
(hereafter J1131-0204 and J1108+0659, respectively) were classified as 
candidate binary AGNs by  \citet{2010ApJ...708..427L, 2010ApJ...715L..30L}, 
on the basis of
spatially resolved double nuclei in the near infrared (NIR) images, whose 
locations are
coincident with the two  velocity components of the narrow line emission 
in the slit spectra \citep{2010ApJ...708..427L}.

Deep NIR images revealed tidal features and double stellar components
separated by $\lsim 2$ kpc in both galaxies \citep{2010ApJ...715L..30L, 
2011ApJ...733..103F}.
In each of the two objects, optical slit 
spectroscopy showed two Seyfert-2 like nuclei that were spatially coincident
with the stellar components (within 1 arcsec) and with velocity offsets
of a few hundred km\,s$^{-1}$ \citep{2010ApJ...708..427L}. On the other 
 hand, for
J1108+0659 integral field spectroscopy showed that the extended narrow line 
region producing the double-peaked profile is far more complex than originally
thought from the longslit spectra, with most of the emission line gas in a 
region oriented perpendicular to the position angle of the two stellar nuclei
\citep{2012ApJ...745...67F}.
Finally, the two galaxies were imaged with Chandra and HST 
\citep{ 2013ApJ...762..110L}. In J1108+0659 both nuclei were detected in 
X-rays
(albeit the south-east nucleus only at $3.4\sigma$ level), while no significant X-ray
emission was detected in J1131-0204.

In this paper we report on Very Long Baseline Interferometry (VLBI) radio
observations at 5\,GHz using the European VLBI Network (EVN) for 
J1108+0659 and J1131-0204, two 
of the four binary AGN classified by \citet{2010ApJ...715L..30L}.
For one source, J1108+0659, our analysis is supported by multi-frequency
observations carried out by the Karl G. Janski Very Large Array (VLA).
For both objects, HST archive observations are available and are used
to complement the discussion.

Throughout, we adopt the WMAP concordance $\Lambda$CDM cosmology with 
$H_0 = 70$~km~s$^{-1}$~Mpc$^{-1}$, $\Omega_{\rm matter} = 0.27$ and 
$\Omega_{\rm vacuum}= 0.73$ \citep{2011ApJS..192...18K}.

\section{Observations and data analysis}
\label{sec,observations}

\subsection{EVN observations}
We observed the radio sources J1131-0204 and J1108+0659 with the
European VLBI Network (EVN) at 5\,GHz (project code EB050). The observations 
were conducted in e-VLBI mode \citep{2008evn..confE..40S} where the data 
stream is not recorded
on disk and later correlated, but is directly transferred to the correlator
through fibre links and processed in real time. The maximum data transmission
rate per station was 1024 Mbps for a total bandwidth of 128 MHz in both left 
and right polarization, using two-bit sampling.
The observations were carried out on 2011 March 22  in 
phase-reference mode \citep[e.g.][]{1995ASPC...82..327B}. 
The nearby calibrators J1136-0330 (at a distance of $1.9\degs$ from 
J1131-0204) and J1112+0724 (at a distance of $0.9\degs$ from 
J1108+0659) were
alternated with the target sources, about 1.5 minutes for the calibrator and 3.5
minutes for the target for a total observing time of about 2 hr on each
target source.
The following radio telescopes provided useful data:
Effelsberg (Germany), Jodrell Bank Mk2 (UK), Medicina (Italy), Onsala (Sweden),
Torun (Poland), Yebes (Spain), Westerbork (The Netherlands), Hartebeesthoek
(South Africa), and Sheshan (China).

\subsubsection{Data calibration and imaging}
\label{sec,cal}

We analysed the correlated data sets using the NRAO Astronomical Image
Processing System ({\it AIPS}; http://www.aips.nrao.edu) for
calibration and imaging purposes. We calibrated the visibility
amplitudes using the system temperatures and gain information provided
for each telescope. We then performed a channel-based inspection and
editing of the data, and corrected the bandpasses using scans on the
bright calibrator 4C39.25.  We also applied standard corrections to
the phases of the sources in our experiment, including ionosphere
corrections (using total electron content measurements publicly
available).  We corrected the instrumental phase and delay offsets
among the baseband converters of each antenna using observations of
the calibrators.

We imaged and self-calibrated (hybrid mapping procedure) the
phase-reference calibrator of each of our target sources, so as to
derive the overall amplitude and phase corrections. We then
transferred those corrections, scan by scan, to the target
sources. Finally, we repeated the fringe-fitting for the
phase-reference calibrators, taking their clean component
models into account to correct for small residual phases resulting from their
structure.  We then interpolated the obtained solutions and applied
them to the target sources. We imaged the calibrated visibilities
of J1131-0204 and J1108+0659 within {\sc
  AIPS}.  We note that we did not perform any self-calibration of the
phases, of the amplitudes, or of the target sources, since the peaks
in emission were too faint for such procedures to be applied.

We point out that we kept the averaging integration time to 1\,s and
used a maximum channel bandwidth in the imaging of 16\,MHz,
which results in a maximum degradation of the peak response for
components far away from the phase centre of less than 2\% at 500 mas
 and prevents artificial smearing of the images
\citep[e.g.][]{1999ASPC..180..371B}.

In imaging our target sources, we applied the following procedures.
When available from high resolution VLA radio images, e.g.  
J1108+0659, we used the information on the position of the
radio components as a prior to search for radio emission on mas
scales. For J1131-0204, neither radio images in the
literature nor observations in the VLA archive were available. We
therefore used the information on the position of the nuclei derived
from HST observations \citep{2013ApJ...762..110L}.
For both sources, in
addition to the fields around the positions of our priors, we imaged
25 additional fields centred on offset positions (where no radio
emission would be expected) to test the reliability of a possible
detection.

\subsection{VLA observations of J1108+0659}

J1108+0659 was observed in A-array configuration at L and
C-bands (on 2011 July 4) and at X-bands (on 2011 July 13) under
project 11A-175 (PI: Fu). The total bandwidth of the observations was 256
 MHz, split in two adjacent IFs centred at 1.4 GHz (L-band), 5.0 GHz
(C-band), and 8.5 GHz (X band). The source 3C\,286 was used as primary flux
calibrator, and the compact radio source J1058+0133 as phase calibrator.
The target source J1108+0659 was observed in L, C, and X bands for about 18, 27, 
and 21 minutes, respectively.
We reduced all three datasets within {\sc AIPS} using standard procedures.

\subsection{HST archival observations}
We also used HST archival observations of J1131-0204 and
J1108+0659. The two sources were observed using the Wide Field
Camera 3 (WFC3) on board the Hubble Space Telescope (HST), programme code GO 
12363. The two targets were imaged in the UVIS/F336W $U$-band  
and IR/F105W (wide $Y$-band) filters. Details of the observations can be 
found in \citet{2013ApJ...762..110L}.
We registered the HST astrometry derived from \citet{2013ApJ...762..110L} using
the images derived from the VLA data of J1108+0659.
We found that the positions derived by \citet{2013ApJ...762..110L} in the $Y$ band are consistent with the position of the radio component C at 5.0\,GHz.
Moreover, with the astrometry derived from \citet{2013ApJ...762..110L} 
the $5\sigma$ radio source detected at 1.4\,GHz at about 9 arcsec SW of 
J1108+0659 overlaps with an optical galaxy in the $Y$-band image.
The absolute astrometric uncertainties of the registered $Y$ and $U$-band images
are $\simeq 0.20\asec$ and $\simeq 0.25\asec$, respectively.

\section{Results and discussion}

\subsection{J1131-0204}
\subsubsection{EVN observations}
\begin{figure} \centering
 \includegraphics[width=80mm,angle=0]{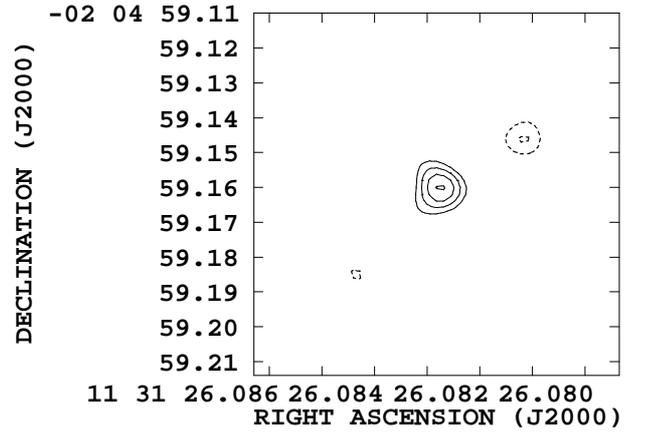} 
 \caption{EVN image at 5\,GHz of J1131-0204 restored with a beam 
size of $15\times 15$ mas. Peak brightness is 118 $\mu$Jy/beam and r.m.s.
noise $\sigma$ is 17 $\mu$Jy. Contour levels are drawn at 
($-4$,$-3$, 3, 4, 5, 6)$\times \sigma$.
}
 \label{evn-image}
\end{figure}

Since no high resolution radio image is available for this source, we
used the position of the two optical nuclei derived from observations
carried out by WFC3 on the HST in the F103W filter \citep{2013ApJ...762..110L}.
Using our 5\,GHz EVN observations we imaged two 
$0\asec.25\times 0\asec.25$ fields around these positions. 
We detected a compact feature in the field centred on
the eastern nucleus, with a signal-to-noise ratio (S/N) of about 6.9
(Fig.\ref{evn-image}). Our EVN image has a resolution of $15\times 15$
mas, a peak surface brightness, and a total
flux density since the source is unresolved,  of 118
$\mu$Jy/beam. 
In contrast, we
detected no radio emission above a $5\sigma$ threshold of 85 $\mu$Jy/beam
in the field centred on the western nucleus (even imaging
larger field sizes). 

Thus, the position of the putative radio core is
$\alpha_{\rm J2000}=$ 11:31:26.088
$\delta_{\rm J2000}$$=-$02:04:59.160,
which is offset with respect to the HST position by less than 0.1
arcsec, so well within the uncertainty of the HST absolute astrometry
\citep{2013ApJ...762..110L}.  
To test the reliability of this radio detection, we imaged 25
additional fields centred at positions that are slightly offset with respect
to the first two in order to be used as test blank fields. In each
field we searched for the brightness peak in a circular region of
radius 0.11 arcsec. We found no brightness peak with S/N$\ge
5.0$ in any of the 25 fields. The mean peak brightness is $60.4\pm
1.7 \mu$Jy/beam
where the error is the error on the mean.  Therefore, we conclude that
the EVN detection of the eastern nucleus of J1131-0204
is significant, whilst we can place only an upper limit for the
western  nucleus.

\subsubsection{The AGN in J1131-0204}
This kpc-scale candidate binary AGN is embedded in a massive disk
galaxy at $z=0.1463$. Two nuclei, separated by 0.6 arcsec,
were detected by NIR imaging \citep{2010ApJ...715L..30L, 2011ApJ...735...48S}. 
The spatial correspondence between the two stellar continuum peaks and the two
[OIII] velocity peaks supported the interpretations that these are two
active galaxies \citep{2010ApJ...715L..30L}. On the basis of line diagnostics,
both nuclei were optically classified as Type-2 Seyferts, implying
that at least one AGN must be present, yet no X-ray emission is
detected at the position of the two nuclei above a 3-$\sigma$ limit, 
implying an upper limit on the (0.5--10 keV) X-ray luminosity of
$< 10^{41}$ erg\,s$^{-1}$ \citep{2013ApJ...762..110L}.

From our EVN observations we derive a radio luminosity at 5\,GHz for the 
eastern AGN of
$L_R=6.5\times 10^{28}$ erg\,s$^{-1}$\,Hz$^{-1}$.
For the western component, assuming as upper limit in the flux density
the mean peak brightness of 60.4 $\mu$Jy, gives $L_R < 3.5\times 10^{28}$
erg\,s$^{-1}$\,Hz$^{-1}$.
We note that these monochromatic radio luminosities are typical of low 
luminosity AGN \citep[e.g.][]{2005A&A...435..521N}.  
The radio core detected at 5\,GHz is
unresolved.  Assuming a limit on the source sizes given by
the FWHM size of the observing beam,
we derive a limit on the brightness temperature
given by $T_b\gsim  10^6$ K.

We notice that neither of the two possible nuclei of J1131-0204 is detected 
in X-rays. \citet{2013ApJ...762..110L} obtained
just an upper limit for the $0.5-10$ keV luminosity of $10^{41}$
erg\,s$^{-1}$. 
We estimated the hard X-ray luminosities for both optical nuclei
expected on the basis of the X-ray to [OIII] luminosity relation for
type 2 AGN reported by \citet{2015ApJ...815....1U}. 
We derived $L_{2-10 kev}\simeq
5.8\times 10^{43}$ erg\,s$^{-1}$ and $\simeq 4.8\times 10^{43}$ erg\,s$^{-1}$
for the candidate western and eastern nuclei, respectively.
The comparison of these estimates with the upper limits inferred by
the Chandra observation implies a mismatch of a factor of $ > 100$. This
can only be explained by the presence of a very dense,
Compton-thick ($N_H \gg 10^{24}$ cm$^{-2}$)  obscuring screen along our line of 
sight to the nucleus. Indeed, a ratio between the observed and intrinsic X-ray
luminosity of a factor of $>70-100$ is typically derived for
Compton-thick Seyfert galaxies \citep[e.g.][]{2001ApJ...563...34M}.
As a merging system, it is likely that a lot of gas and dust have been
funnelled into the nuclear region, providing a likely explanation
for the apparent buried nature of the AGN. 

Using the definition for the radio-loudness parameter
of $R_X=\nu L_\nu(\rm 5 GHz)/L_X$  \citep{2003ApJ...583..145T}, we
derived $\log(R_X)\simeq - 5.2$ for the radio detected eastern nucleus,
assuming the hard X-ray luminosity derived from the [OIII] emission.
This value characterizes the eastern nucleus of J1131-0204
as a radio-quiet AGN.

Summarizing, from our EVN observations we obtain evidence of just one
radio-quiet AGN in J1131-0204, whose position is coincident, 
within the errors, with the eastern optical nucleus.

\subsection{J1108+0659}

\subsubsection{VLA observations}
\begin{figure} \centering
 \includegraphics[width=70mm,angle=0]{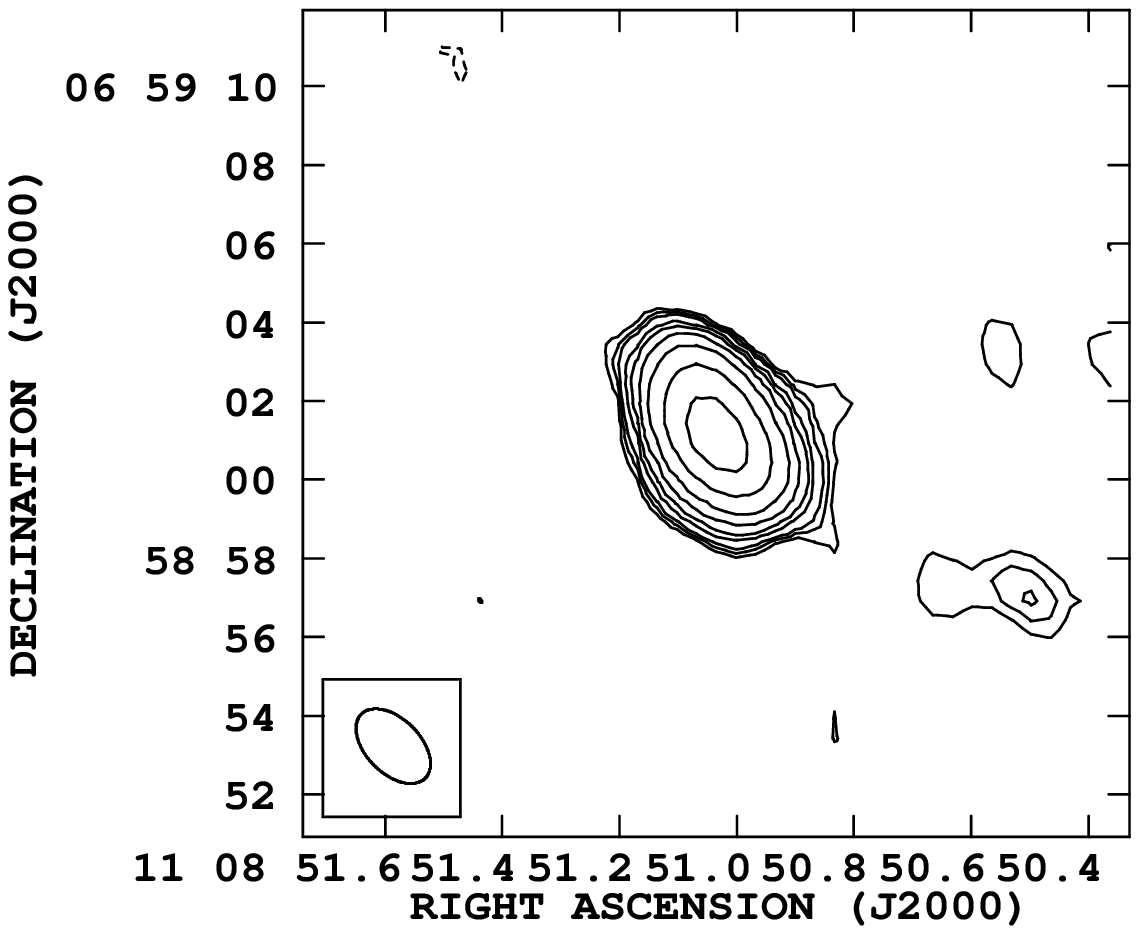} \\
 \includegraphics[width=70mm,angle=0]{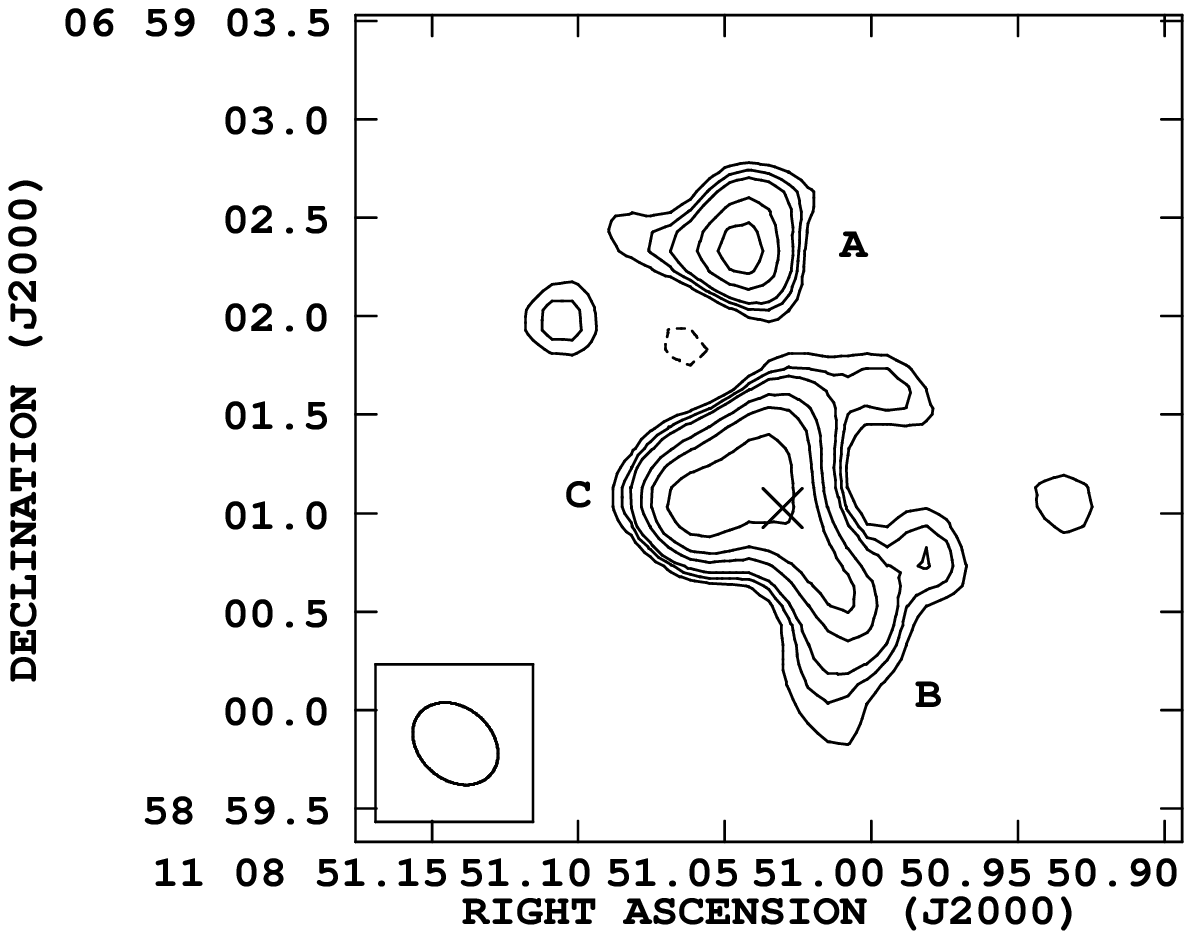} \\
 \includegraphics[width=70mm,angle=0]{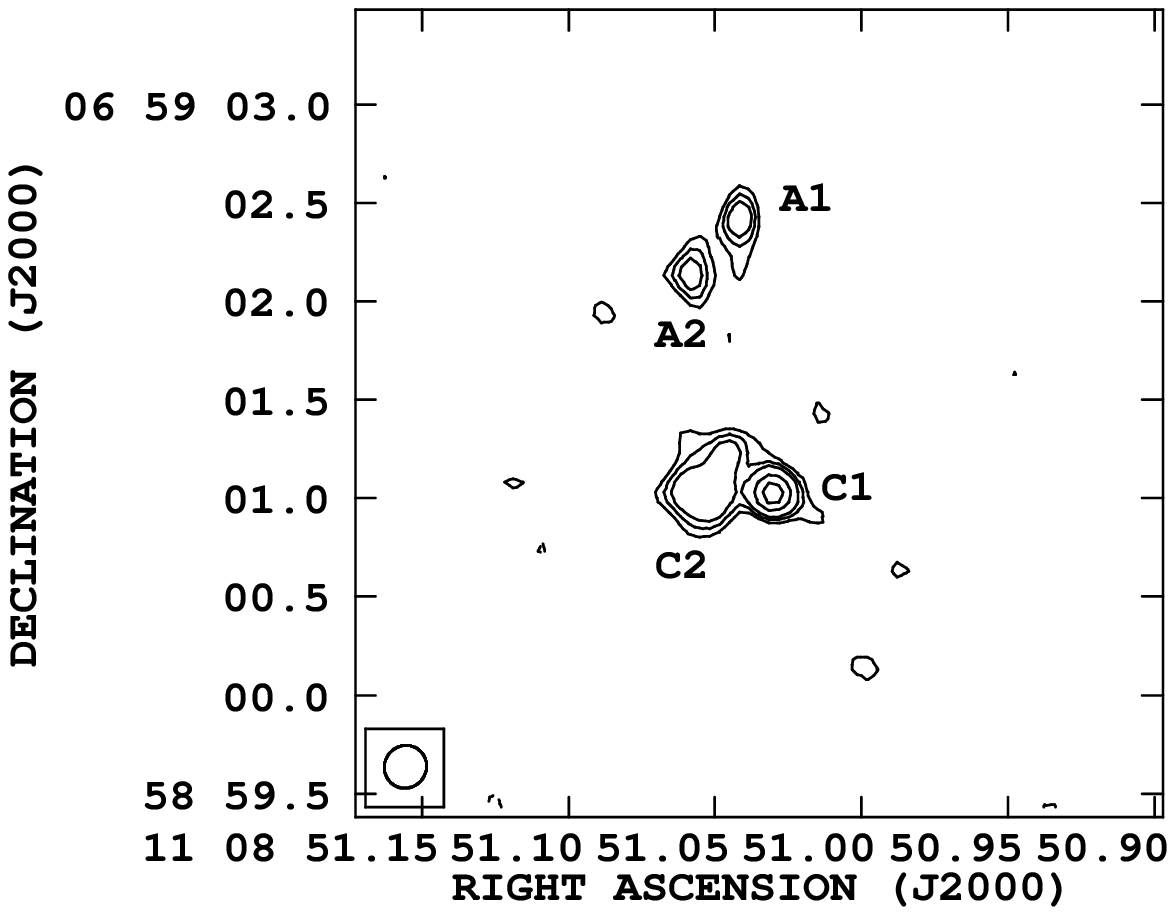} 
 \caption{VLA images of J1108+0659. {\it Top panel}: image at
1.4\,GHz, beam size is $2.30\times 1.38$ arcsec, peak flux is 5.82 mJy/beam, and
r.m.s. noise $\sigma=0.040$ mJy. {\it Middle panel}: image at
5.0\,GHz, beam size is $0.46\times 0.37$ arcsec, peak flux is 0.41 mJy/beam,  
r.m.s. noise $\sigma=0.020$ mJy. {\it Bottom panel}: image at
8.5\,GHz, beam size is $0.22\times 0.21$ arcsec, peak flux is 0.20 mJy/beam,
r.m.s noise $\sigma=0.017$ mJy. For all three images, contour levels are 
($-3,3,4,5,7.5,10,15,25,50,100$)$\times\sigma$. Labels identify components 
discussed in the text, the cross in the middle panel is drawn at the position of component C1.}

 \label{vla-image}
\end{figure}

In Fig.\,\ref{vla-image} we
display the contour maps of the images obtained by the VLA observations at
 1.4, 5.0 e 8,5 GHz.
At 1.4\,GHz (top panel), the source is slightly resolved with a resolution of $2.30\times 1.38$
arcsec, and is fitted by a Gaussian
component with a deconvolved FWHM size of $1.9\times 1.0$ arcsec in
position angle $16\degs$.  
The fitted total flux is $9.82\pm 0.15$ mJy, consistent
with the flux of 9.84 mJy measured in the FIRST survey 
\citep{1997ApJ...475..479W}.

At sub-arcsecond resolution, the 5.0\,GHz and 8.5\,GHz images
(middle and bottom panels in Fig.\,\ref{vla-image}) show interesting
and new features. At the intermediate frequency (resolution
$0.46\times 0.37$ arcsec), the source morphology is mainly
characterized by three components.  The brightest central component 
(labelled as C) is
elongated at position angle $113\degs$ and can be fit either by a single
extended component or by two more compact components embedded in some
diffuse emission, mainly on the north-western side.  The total flux of
component C is about 1.4 mJy.  A tail of extended emission
(labelled as B)
is detected in the south-western direction from the central
component. This emission extends for more than 1 arcsec from 
component C and accounts for $\sim 0.5$ mJy.  On the northern
side, a resolved component (deconvolved FWHM size of $0.38\times 0.25$
arcsec in position angle $133\degs$) with a total flux density of 0.38
mJy is detected (labelled A). Accounting for the extended emission between the
three major components, the total flux density at 5.0\,GHz is
$2.6\pm 0.1$ mJy.

At the highest resolution ($0.22\times 0.21$ arcsec) of the 8.5\,GHz
image, the extended emission of the south-western tail is resolved out.
The central component C is resolved into a  compact
core, labelled C1, detected at position $\alpha_{J2000}=$11:08:51.030 
$\delta_{J2000}=$06:59:01.03 and a resolved component C2. 
The core C1 is only slightly
resolved with a fitted FWHM size of $0.14\times 0.04$ arcsec in position angle
$52\degs$. The northern component detected at 5.0\,GHz is resolved in two
components, A1 and A2, separated by 0.39 arcsec (about 1.2 kpc) in position 
angle $140\degs$, 
consistent with the single component fit at 5.0\,GHz. The two components
A1 and A2 are  both slightly resolved in the NS direction.

\subsubsection{EVN observations}
We used the information on the position of the radio components A1, A2,
and C1 from the 8.5\,GHz VLA image as a prior to search for
radio emission in our EVN observations.  We detected no radio emission
associated to any of these components above a significant level of
confidence ($\ge 5\sigma$).  We therefore put a limit on the flux density
of the mas-scale core of $\sim$ 85 $\mu$Jy at 5\,GHz.  
There are obvious possible reasons for
not having detected C1, the most compact VLA component in our EVN 
observations.  It is clear
from the radio morphology and fitted size that, at the resolution of
the VLA, the component C1 can still be affected by emission
from the eastern lobe, C2, and/or from a jet component feeding the lobe on
scales of several tens of mas.  Therefore it is possible that we did
not detect a core component on the mas scale due to the sensitivity
limit of our EVN observations.  Observations filling the gap in
resolution between the 8.5\,GHz VLA data and our EVN observations at
5\,GHz would help to solve this issue.

\subsubsection{Radio spectral index}
\begin{figure*} \centering
 \includegraphics[width=80mm,angle=0]{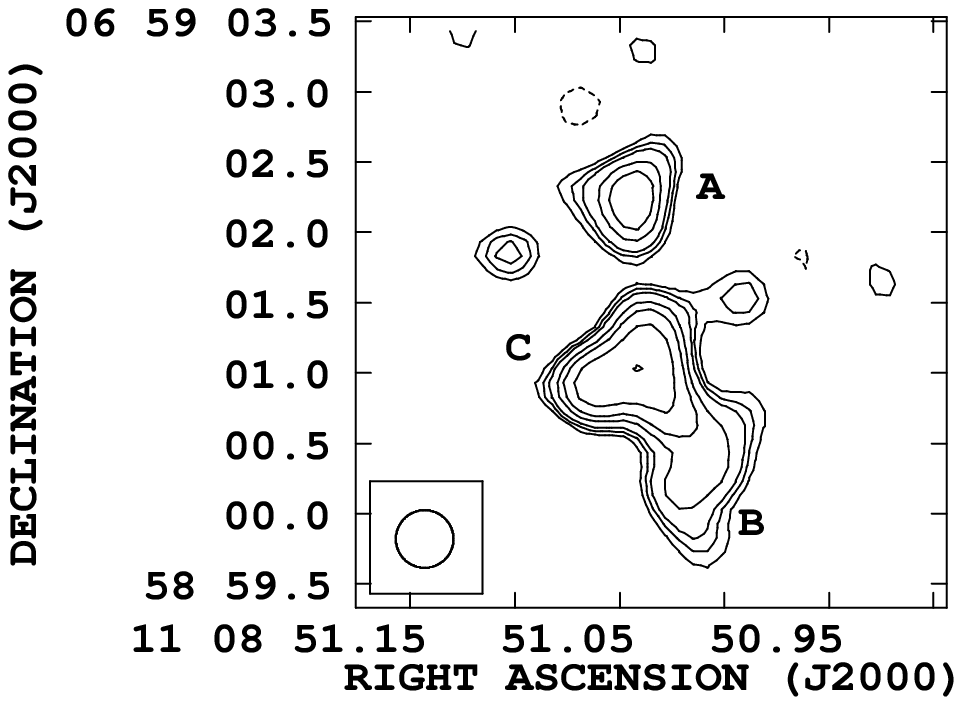}
 \includegraphics[width=80mm,angle=0]{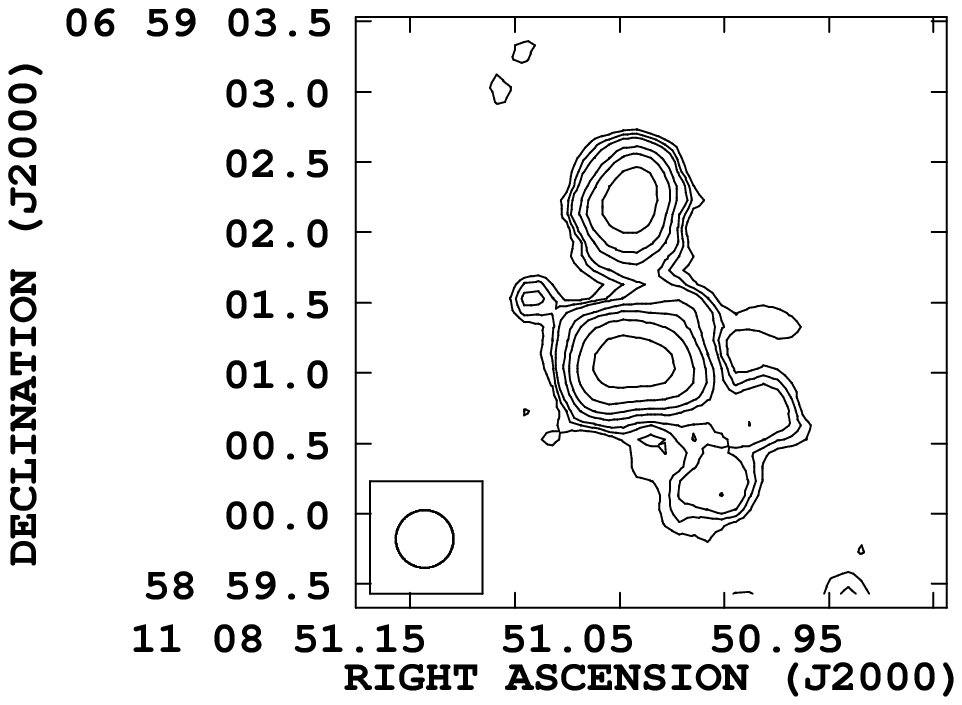}
 \caption{VLA images of J1108+0659 used for spectral index
analysis. {\it Left panel}: image at
5.0\,GHz, beam size is $0.41\times 0.41$ arcsec, peak flux is 0.50 mJy/beam,  
r.m.s noise $\sigma=0.025$ mJy. {\it Right panel}: image at
8.5\,GHz, beam size is $0.41\times 0.41$ arcsec, peak flux is 0.34 mJy/beam,
r.m.s noise $\sigma=0.015$ mJy. For all three images, contour levels are 
$-3,3,4,5,7.5,10,15,25,50,100\times \sigma$.}
 \label{vla-spix-images}
\end{figure*}

We derived radio spectral index information for J1108+0659 using the VLA 
data at 5.0\,GHz and
8.5\,GHz, as the resolution of the 1.4\,GHz image is too poor
to derive the spectral index of different radio
components.  To this end, we imaged the two data sets with the same
resolution ($0.41\times 0.41$ arcsec; see Fig.\ref{vla-spix-images}).

Then, we combined the images at the two frequencies to produce a
spectral index image and a spectral index error image.
We identified three main components for which we extract the spectral
index information presented in Table \ref{tab,spix}. 
For each
component we list the mean spectral index (the mean of the spectral 
index values over the region covered by the component) and the dispersion
of the spectral index values with respect to the mean 
($\sigma$) and the mean error, i.e. the mean from the spectral index error
image in the region ($\sigma_M$).
\begin{table}
\centering                          
\caption{Mean radio spectral index, $\alpha$ ($S_\nu \propto
  \nu^{-\alpha}$) of selected regions in J1108+0659. $\sigma$ is the dipersion
of $\alpha$, and $\sigma_M$ is the average error on $\alpha$.}             
\label{tab,spix}      
\renewcommand{\footnoterule}{}
\begin{tabular}{llllll}
\hline \hline
 Comp. & $\alpha$  & $\sigma$ & $\sigma_M$ \\
      &       &           & \\
\hline 
A & 0.4   & 0.07 & 0.2 \\ 
C   & 0.8    & 0.5  & 0.2 \\ 
B   & 1.7    & 0.8  & 0.5 \\ 

\hline
\end{tabular}   
\end{table}
The northernmost radio component (A) contains both components A1 and A2,
which cannot be resolved at this resolution. The spectral index in
this region is very smooth and uniform ($\sigma=0.07$) throughout the
region, implying that both A1 and A2 have similar spectral
properties. The mean spectral index of this component is $0.4\pm 0.2$,
where the error is the mean error of the derived spectral index values
in this region.  This radio component is located in the northern tail
of star formation activity. 

The central component (C) is clearly extended at both frequencies
(Fig.  \ref{vla-spix-images}). Its is worth noting 
how the shape, in particular the position angle, of this component 
changes with  frequency.  At 5.0\,GHz, the component is extended 
with about the same position angle of the separation between the two 
optical nuclei and the position angle of the extended $U$-band emission
starting from the north-west optical nucleus that traces recent star
formation (Fig. \ref{vla-hst}).
We also note how the 5.0\,GHz radio emission follows
a filament of star formation activity in the north-west region of this
component.

At higher frequency,  the position angle of component C flips by about 
$40\degs$, most likely due to the emergence of a component with a 
flatter spectrum. This emerging component is  the candidate radio nucleus 
C1, and it becomes the brightest component in the high resolution
8.5\,GHz image (see Fig.\ref{vla-image}). The average spectral index in region
C is steep,  but not homogenous, and the rather large dispersion ($\sigma=0.5$)
is due to large variations in the spectral index across the component,
owing to the superposition of components with different spectral shapes.

The southern component B has a very steep spectral index $\alpha
=1.5$ with $\sigma=0.8$. In this region the large dispersion is mainly due to
the large errors associated to the spectral index values.
The radio emission is clearly associated to the 
extended $U$-band emission-tracing region of star formation activity.
Finally, we made an attempt to estimate the spectral index of component C1.
Component C1
has a total flux of about 200 $\mu$Jy and a deconvolved FWHM of
$0.14\times 0.04$ arcsec in position angle $52\degs$.  This component
is not prominent in the 5.0\,GHz image, since it is embedded in
the more diffuse emission of component C. This makes it challenging to
derive the radio spectral index of this component.  We estimated the radio
spectral index of C1 using flux densities extracted from the
same region at 8.5\,GHz and 5.0\,GHz using images at the highest
(but different) resolutions, and obtained
$\alpha_{C1}\simeq 0.3\pm 0.2$.  The uncertainty in the spectral
index value is due to the estimate of the contribution of the diffuse
emission to the flux density measured at 5.0\,GHz.

\subsubsection{AGN and/or star formation emission in J1108+0659}
Near-infrared imaging revealed
two stellar nuclei, which are separated by $\simeq 0.7$ arcsec in position angle 
$141\degs$and completely enshrouded by the extended narrow line emission  
 \citep{2010ApJ...715L..30L, 2011ApJ...733..103F, 2013ApJ...762..110L}.  
The extended narrow line region (ENLR) is centred
on the double stellar components and extends for several arcseconds in
position angle $\theta_{ENLR}\simeq 0\degs$ \citep{2012ApJ...745...67F}. 
All the line
diagnostics are consistent with the presence and dominance of at least
one type-2 AGN with Seyfert-like luminosity \citep{2010ApJ...708..427L,
2010ApJ...715L..30L, 2011ApJ...733..103F, 2012ApJ...745...67F,
2012MNRAS.421.1043S}. Although with limited
confidence, both stellar nuclei are detected in the X-ray Chandra
observations \citep{2013ApJ...762..110L}. 
The hardness ratio derived for the north-west nucleus is
consistent with a steep spectral index of $\Gamma\simeq 2.4$, which is
typically associated to starburst galaxies \citep{1998ApJS..118..401D}.
This hypothesis is also 
supported by the estimated soft/hard X-ray luminosity of $4\times 10^{41}$
erg\,s$^{-1}$, which is in the starburst range.
In contrast, the south-east nucleus exhibits both the X-ray flat spectral shape
and luminosity of an obscured AGN, although it was detected at a
$3.4\sigma$ level. Specifically, \citet{2013ApJ...762..110L} 
measured a column
density of $N_H\simeq 3\times 10^{22}$ cm$^{-2}$, which implies an X-ray
de-absorbed luminosity of $\simeq 1.2\times 10^{42}$ erg\,s$^{-1}$,
suggesting the presence of an AGN.

On the other
hand, the UV-continuum is centred on the north-west stellar component and
extends for a few arcseconds along the same position angle of the
[OIII] line emission.
Using the X-ray to [OIII] luminosity relation for
type 2 AGN reported by \citet{2015ApJ...815....1U}, we derive  hard X-ray 
luminosities
of $1\times 10^{44}$ erg\,s$^{-1}$ and $1.6\times 10^{43}$ erg\,s$^{-1}$
for the north-west
and south.east nucleus, respectively. While the value for the south-east nucleus
is broadly consistent with that meaured by \citet{2013ApJ...762..110L}, there 
is a mismatch of a factor $\sim 100$ for the north-west component.

\begin{figure} \centering
 \includegraphics[width=87mm,angle=0]{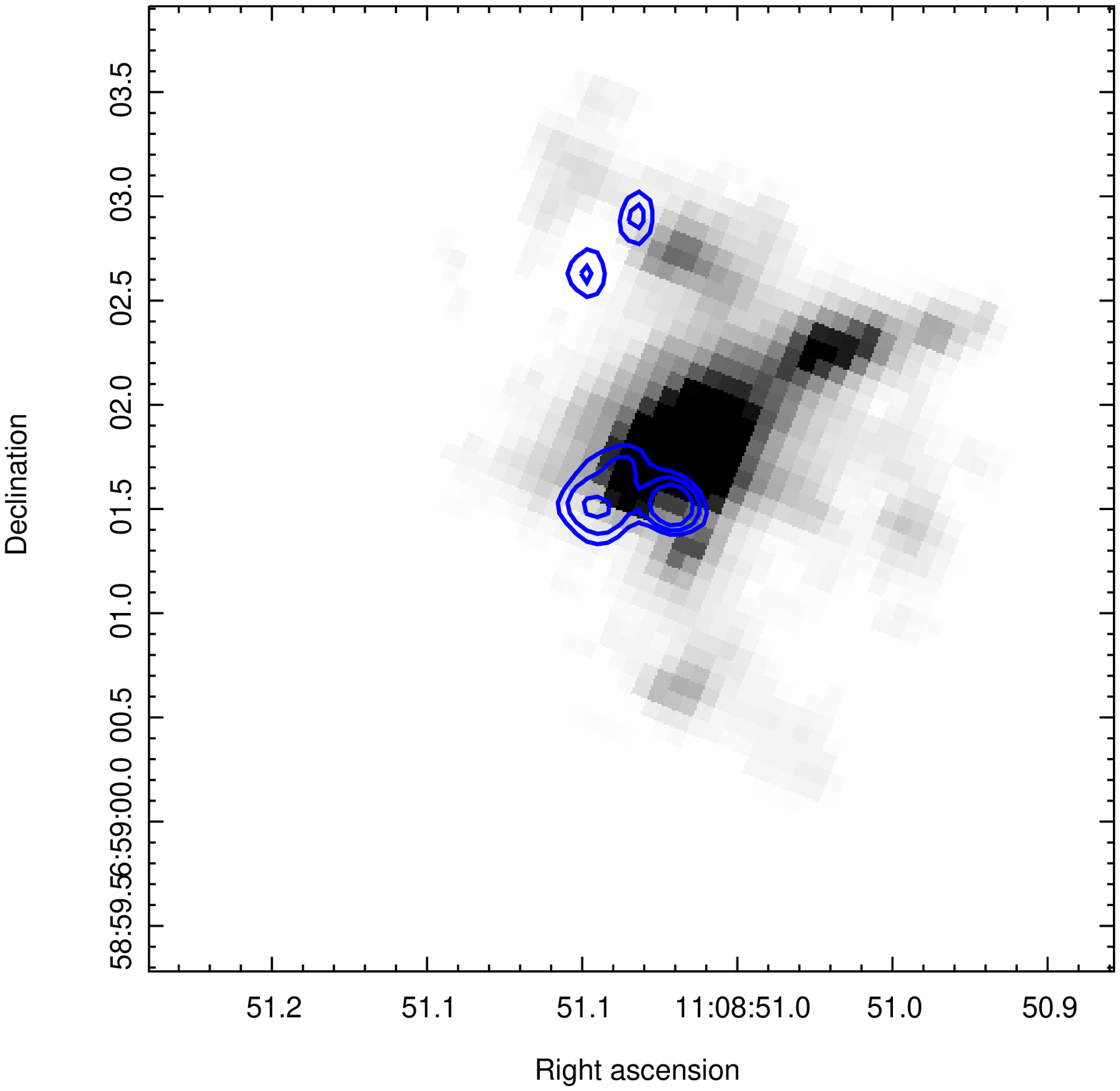}
 \includegraphics[width=80mm,angle=0]{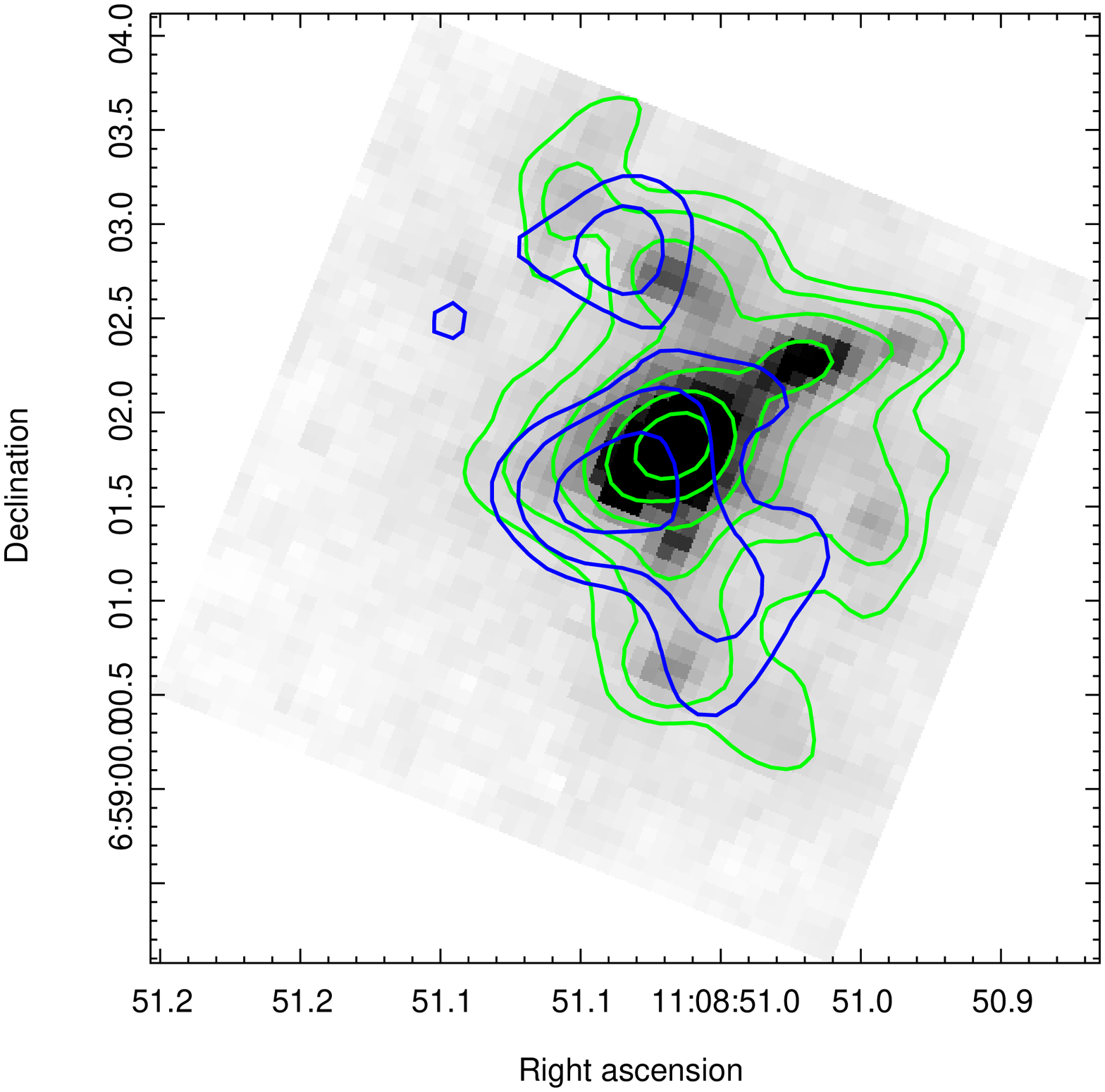}
 \caption{HST U-band image of J1108+0659 
overplot with the 8.5\,GHz VLA image (upper panel, HST in greys and VLA in 
contours)  and with the 5.0\,GHz VLA image (lower panel, HST in greys and green
contours and VLA in blue contours). 
The overplot with the 8.5\,GHz radio image shows the slight offset between
the candidate radio core and the north-west core detected in the UV continuum. The south-east
core is undetected in the UV continuum.
The overplot with the 5.0\,GHz radio image clearly shows that the extended
radio emission and the UV continuum are cospatial.
}
 \label{vla-hst}
\end{figure}

C1, which is the most compact radio component and has the flattest
spectral index, is the best candidate for hosting the radio
nucleus of J1108+0659, although the offset of the position of C1
with respect to the HST and X-ray position of the north-west
optical/NIR core ($0\asec.29$ and $0\asec.23$, respectively)
may cast some doubts. 
Still, this value is comparable to the uncertainty in the HST
astrometry. In addition, the relative alignment of the radio extended
emission with the HST $U$-band emission is rather good, and it would be
worsened if the north-west optical core would be shifted to the position of
C1. Figure~\ref{vla-hst} shows the offset between component
C1 and the north-west UV/NIR core.
Components A1 and A2
have a slightly steeper spectral index than C1, are less compact,
and are distant ($>1$ arcsec) from both  stellar optical/NIR nuclei, and
therefore they do not seem to be good candidates for hosting the radio core 
of an AGN. Moreover, they are located at the northern peak of
the [OIII] narrow line emission \citep[see Fig.3 in][]{2012ApJ...745...67F}, 
in a region where intense star formation is occurring.

Summarising, from our VLA observations, we find some evidences  that 
component C1 is the radio nucleus of the AGN: it has the flattest radio
spectrum and is the most compact component in our highest frequency image.
On the other hand, we note that the presence of a mas-scale
compact core is not confirmed by our EVN observations, albeit this could 
be explained by lack of sensitivity, and that the position of
component C1 is offset with respect to the HST and
X-ray Chandra position of the eastern core.  
Further high sensitivity data, with an angular resolution filling the gap 
between the 8.5\,GHz VLA and 5\,GHz EVN observations, are needed to 
unambiguously 
confirm the presence of a radio AGN in J1108+0659.

The extended UV emission is not the only evidence for intense star 
formation in J1108+0659.
The source is associated to the IRAS Faint Source Catalog galaxy
F11062+0715 \citep{1992ifss.book.....M, 2009ApJ...704..789H}. 
While the association
has been claimed [by others] as tentative due to the large offset between the
IRAS and optical positions ($\simeq$9 arcseconds), we note 
that the $1\sigma$ ellipse error for F11062+0715 is $27\times 7$
arcsec, and therefore the association seems secure within the
uncertainties of the IRAS position.  F11062+0715 has ULIRG-like
luminosity with a total IR luminosity
L$_{\rm IR}(1-1000\, \mu m)=12.01$ \citep{2009ApJ...704..789H}.

An overplot of the radio contours at 5.0\,GHz over the
$U$-band HST image is shown in Fig.\,\ref{vla-hst}. The $U$-band image reveals
intense star formation activity in the north-west nuclear region, while the 
south-east nucleus
is undetected, indicating a very low level of star formation around this 
nucleus, or that the star formation is highly obscured 
\citep{2013ApJ...762..110L}.

The extended radio emission is co-spatial with the $U$-band UV
emission and with the [OIII] line emission, supporting  
intense star formation as a common origin for the radio and UV continuum. 
We notice that the different indicators
of star formation rate (SFR) give different results. Indeed, on the basis of the
$U$-band luminosity, \citet{2013ApJ...762..110L} derived a star formation rate
SFR$_{UV}\simeq 10$ \msunyr. On the other hand, using the total
IR luminosity yields a SFR$_{IR}\simeq 100$ \msunyr, an order
of magnitude greater than derived from the $U$-band emission.  To
derive the star formation rate
from the radio emission, we need to separate the
synchrotron emission related to the AGN (core and lobe/jet emission)
from what is associated with star formation activity. For simplicity, we
assume here that component C (including C1) is produced by
AGN-related radio emission. The total flux density at 5.0\,GHz of
components A and B is about 0.9 mJy. Using an average spectral index
$\alpha=0.8,$ we obtain a radio luminosity at 1.4\,GHz associated to
star formation $L_{SF}\simeq 2.2\times 10^{30}$ erg\,s$^{-1}$\,Hz$^{-1}$
corresponding to a SFR$_{1.4GHz}\simeq 110$ \msunyr
consistent with the rate derived from the total infrared luminosity.  We
conclude that a large part of the UV-emission must be obscured,
yielding a SFR estimate in this band an order of magnitude less than derived from measurements unaffected by dust absorption.

\section{Summary}
We have reported on EVN observations at 5\,GHz of two dual AGN candidates
J1131-0204 and J1108+0659 (supported by multifrequency VLA observations 
for this source).
The main goal of these observations was to detect, in both objects,
two radio cores to unambigously confirm the presence of a dual AGN.
Our results can be summarized as follows:

\begin{itemize}
\item J1131-0204. We detect only one compact radio core, 
associated
with the eastern optical/NIR nucleus of this system with a 5\,GHz luminosity
of $L_R=6.5\times 10^{28}$ erg\,s$^{-1}$\,Hz$^{-1}$ typical of low luminosity
AGN. For the second possible AGN in this system we can only provide a $5\sigma$ 
upper limit on its 5\,GHz radio luminosity of 
$L_R=3.5\times 10^{28}$ erg\,s$^{-1}$\,Hz$^{-1}$.
No X-ray emission is detected in either of the optical/NIR nuclei,
even though the X-ray to [OIII] luminosity relation for type 2 AGN would yield  $L_{2-10 kev}\sim 10^{43}$ erg\,s$^{-1}$ for each of the muclei. 
This is a clear
indication that the AGN is strongly obscured by a Compton-thick screen.

\item J1108+0659. This merging system is rather puzzling. 
From our EVN observations, we can only put upper limits on the radio luminosity 
of possible compact cores on the milliarcsecond scale. 
Using VLA observations we find a complex 
radio morphology on sub-arcsecond scale with a tentative core
identification at 8.5\,GHz. The core (component C1) has a flat radio spectrum and 
is slightly resolved, and its non detection 
in the EVN observations could be consistent with the sensitivity limit of 
our observations.  Moreover, the position of this tentative radio core is
offset ($\sim 0.2$ arcsec) with respect to the optical/NIR and X-ray
positions of the closest  nucleus.
Further radio observations that fill the gap in angular resolution
between the 8.5\,GHz VLA and the 5\,GHz EVN observations are needed to 
establish the nature of the radio core candidate C1.
Most of the extended sub-arcsecond scale radio emission is closely related to 
the UV continuum and with the [OIII] line emission, suggesting that it could 
come from intense star formation.
Disentangling the AGN-related from star formation-related radio emission,
we derived SFR $\simeq 110$ \msunyr that is consistent with what is expected from 
the total
infrared luminosity, but an order of magnitude higher than implied
by the UV luminosity, in turn implying large obscuration by  dust.

\end{itemize} 

\begin{acknowledgements}

The European VLBI Network is a joint facility of independent European,
African, Asian, and North American radio astronomy institutes. Scientific
results from data presented in this publication are derived from the
following EVN project EB050.

MPT and EP are members of the MAGNA project
(http://www.issibern.ch/teams/agnactivity/Home.html ), and they gratefully
acknowledge support of the International Space Science Institute
(ISSI) in Bern, Switzerland.
MPT acknowledges support by the Spanish MINECO through grants AYA
2009-13036-C02-01 and AYA 2012-38491- C02-02, cofunded with FEDER funds.
MPT also acknowledges the hospitality of the Department of Theoretical
Physics of the University of Zaragoza, where part of this work was done.

The authors thank an anonymous referee for useful suggestions and comments
that have improved the paper.
\end{acknowledgements}
\bibliographystyle{aa}
\bibliography{mb-biblio}

\end{document}